# Robust Anomalous Hall Effect in Epitaxial Cubic $D0_3$-$Fe_3Ge$ Nanoplates Grown by Chemical Vapor Deposition

*Mingxuan Chen*[1,†], *Chao Chen*[1,†], *Siqi Jiang*[1,†], *Shuxiang Qiao*[2,†], *Hongyan Lu*[2], *Di Zhang*[1], *Jiabei Huang*[1], *Fuzhuo Lian*[1], *Zhaochen Qu*[1], *Siqin Wang*[1], *Ziyi Wang*[1], *Shuai Zhang*[1], *Yulu Liu*[1], *Ronghua Liu*[1,3,4], *Xiaoxiang Xi*[1], *Renjun Du*[1] *, *Lei Wang*[1,5,6] *, *Geliang Yu*[1,5,6] *

[1]National Laboratory of Solid State Microstructures, Collaborative Innovation Center of Advanced Microstructures, School of Physics, Nanjing University, Nanjing 210093, China.

[2]School of Physics and Physical Engineering, Qufu Normal University, Qufu 273165, China.

[3]Jiangsu Provincial Key Laboratory for Nanotechnology, Nanjing University, Nanjing 210093, China.

[4]National Key Laboratory of Spintronics, Nanjing University, Suzhou 215000, China.

[5]International Joint Laboratory on Two Dimensional Materials, Nanjing University, Nanjing 210093, China.

[6]Jiangsu Physical Science Research Center, Nanjing University, Nanjing 210000, China.

†These authors contributed equally: Mingxuan Chen, Chao Chen, Siqi Jiang, Shuxiang Qiao.

[*] E-mail: renjundu@nju.edu.cn; leiwang@nju.edu.cn; yugeliang@nju.edu.cn

# Abstract:

The iron-based ferromagnetic materials have emerged as a promising system for spintronics due to their high Curie temperatures and significant anomalous Hall effect (AHE) driven by Berry curvature. While there have been investigations into the epitaxial growth of non-layered iron-based ferromagnetic materials, the realization of high-quality single crystal thin films continues to be a formidable obstacle. Meanwhile, the issues such as grain boundary scattering further hinder their characterization and application. Here, we report the first chemical vapor deposition (CVD) synthesis of single-crystalline cubic $D0_3$-$Fe_3Ge$ nanoplates. This compound represents a substantial departure from the conventional hexagonal phases typically observed in the Fe-Ge system. Structural characterizations confirm the cubic lattice and stoichiometric purity of the crystals. Structural analysis identifies a strong epitaxial effect, resulting in a giant compressive strain of ~4.16% at the interface that relaxes gradually toward the surface. The second-harmonic generation (SHG) signals might further reveal that such interfacial strain disrupted the inherent centro-symmetry of the cubic phase. SQUID magnetometry shows robust room-temperature ferromagnetism and clear magnetic anisotropy. Magneto-transport measurements reveal a robust AHE that remains stable at temperatures far exceeding room temperature. The scaling observed between the anomalous Hall resistance and longitudinal resistance strongly suggests that this effect is dominated by the intrinsic Berry curvature mechanism. Furthermore, density functional theory (DFT) calculations reveal that the huge epitaxial strain induces band reconstructions and creates dense spin-orbit coupling gaps near the Fermi level, effectively doubling the anomalous Hall conductivity to match our experimental observations.

# 1. Introduction:

The exploration of magnetic order in materials with spatially confined geometries is pivotal for the advancement of next-generation spintronic devices, such as high density memory and magnetic sensors.[1–7] Notably, these materials have enabled the induction and manipulation of novel electronic transport phenomena, such as the anomalous Hall effect (AHE), particularly when coupled with strong intrinsic or interface-induced spin–orbit coupling. More recently, research on the AHE has undergone a significant shift, from being regarded primarily as a standard magneto-transport phenomenon to serving as a vital tool for probing topological quantum states and advancing next-generation spintronic devices.[8–23] The intrinsic mechanism of the AHE, originating from the Berry curvature of the electronic band structure, has garnered immense interest because it can lead to large Hall conductivities that are robust against temperature and scattering.[24–26] This intrinsic AHE is closely linked to the topology of the Bloch wavefunctions and is amplified by band-crossing features such as Weyl points, nodal lines, or anti-crossing gapped nodes in the momentum space. Materials hosting such topological singularities near the Fermi level are highly sought after for achieving large anomalous Hall conductivity (AHC). This pursuit has led to the discovery of remarkably large AHC in several material systems, including the Kagome magnet $Co_3Sn_2S_2$,[20,27,28] the ferromagnetic Weyl semimetal $Co_2MnGa$,[29] and the ferromagnetic van der Waals semimetal $Fe_3GeTe_2$.[30] However, many topological magnets with large AHC also exhibit high coercivity, requiring large magnetic fields for magnetization switching, which is detrimental to device integration and energy efficiency. Furthermore, the Curie temperatures ($T_C$) of most such ferromagnetic systems lie below room temperature, severely limiting their practicality in low-dimensional spintronic devices.[31–33] In this context, the bulk iron-germanide $D0_3$-$Fe_3Ge$ exhibits promising properties: a high $T_C$ well above room temperature (exceeding 740 K),[34] a large saturation magnetization (~5.29 μB/f.u.), and low coercivity ($H_C$). Prior investigations into polycrystalline bulk $Fe_3Ge$ have established an intrinsic AHC of 106 $\Omega^{-1}\cdot\mathrm{cm}^{-1}$ for the $D0_3$-$Fe_3Ge$, explicitly attributing its origin to the Berry curvature of the electronic band structure,[35] making it a promising candidate for room-temperature spintronics.

However, in polycrystalline bulk samples, grain boundaries and intrinsic defects act as strong scattering centers, inevitably suppressing charge transport and restricting the experimentally achievable AHC to values well below the theoretical upper bound. Furthermore, these extrinsic scattering contributions obscure a precise analysis of the intrinsic Berry curvature. Therefore, the synthesis of high-quality, single-crystalline thin nanoplates is imperative to unlock the full potential of $Fe_3Ge$'s anomalous Hall properties. Despite its promising attributes, the growth of phase-pure, single-crystalline $D0_3$-$Fe_3Ge$ nanoplates has remained elusive. Conventional physical vapor deposition (PVD) techniques often struggle with stoichiometric control and the avoidance of phase segregation, hindering the exploration of intrinsic transport properties in the clean limit. Because $Fe_3Ge$ is inherently non-layered, its low-dimensional counterparts cannot be realized through conventional mechanical exfoliation. Hence, the bottom-up growth of $Fe_3Ge$ nanoplates has, to date, posed a significant challenge.

To access the intrinsic physics, it is imperative to fabricate defect-free, single-crystalline domains on the mesoscopic scale, which minimizes extrinsic scattering channels. Herein, we overcome this long-standing synthetic barrier by reporting the first successful epitaxial growth of single-crystalline $D0_3$-$Fe_3Ge$ thin nanoplates via chemical vapor deposition (CVD). This growth technique provides superior control over stoichiometry and crystallinity, enabling access to a previously unexplored regime of material quality for this compound. Moreover, our findings show that the CVD-grown $D0_3$-$Fe_3Ge$ high crystalline quality nanoplates exhibit an interfacial strain of ~4.16%. Despite existing studies on the epitaxial growth of $Fe_3Ge$ thin films,[36,37] continuous films typically develop into polycrystalline structures during growth, which contain numerous grain boundaries. The growth of single crystal nanoplates can eliminate such additional scattering and highlight the intrinsic contribution of the Berry curvature. Furthermore, when grown on substrates with a lattice mismatch (e.g., sapphire or mica), continuous films are highly prone to generating defects, such as misfit dislocations, to release stress. In contrast, due to their limited lateral dimensions, nanoplates can better accommodate the substrate through elastic deformation, thereby sustaining a high and uniform epitaxial strain without introducing substantial dislocations. More importantly, these nanoplates possess robust room-temperature ferromagnetism and clear magnetic anisotropy with an in-plane magnetic easy axis. Furthermore, D0$_3$-$Fe_3Ge$ nanoplates demonstrate robust AHE extending well beyond room temperature. The scaling behavior between anomalous Hall resistance and longitudinal resistance ($\log R_{AHE} \propto \log R_{xx}$) consistent with the intrinsic Berry curvature contribution. Supported by density functional theory (DFT) calculations, we demonstrate that this interfacial strain significantly modulates the electronic band structure, which substantially enhances the anomalous Hall conductivity compared to its unstrained bulk counterpart. Our work establishes CVD as a viable route for synthesizing high-quality iron-based intermetallic nanoplates and validates D0$_3$-$Fe_3Ge$ as an exceptional candidate for room-temperature spintronic applications. Exhibiting a robust AHE alongside high $T_C$ and low $H_C$, this material is highly promising for developing advanced devices, such as low-power Hall sensors and low-field magnetic sensors, paving the way for exploring transport phenomena in high-quality epitaxial nanoplates of related intermetallic compounds.

## 2. Results and Discussion

**Synthesis of D0$_3$-$Fe_3Ge$ nanoplates**

Figure 1a depicts a schematic of the CVD process used to grow cubic DO$_3$-$Fe_3Ge$. A mixture of germanium(II) iodide ($GeI_2$) and iron(II) dichloride ($FeCl_2$) powders served as precursors and were positioned directly beneath the sapphire substrates to ensure sufficient vapor pressure. A growth temperature of approximately 600 °C was selected to thermodynamically favor the formation of the D0$_3$-$Fe_3Ge$ phase. During the CVD process, the stoichiometry of the grown nanoplates is dictated by the dynamic flux ratio of the precursor vapors arriving at the substrate. Because $GeI_2$ (melting point 460 °C) possesses a higher saturated vapor pressure and vaporization rate than $FeCl_2$ (melting point 677 °C), a strictly stoichiometric 3:1 precursor mixture would yield a Ge-rich vapor atmosphere, leading to the formation of Ge-rich impurity phases. To overcome this kinetic mismatch, a slight excess of $FeCl_2$ was employed. This excess mass compensates for the slower vaporization rate of $FeCl_2$, dynamically tuning the partial pressures in the carrier gas to ensure that the effective Fe:Ge flux ratio reaching the sapphire substrate remains optimal for the continuous epitaxial growth of the pure $D0_3$-$Fe_3Ge$ phase. Following the growth process, the furnace was allowed to cool naturally to room temperature. As shown in Figure 1c,d, the $D0_3$-$Fe_3Ge$ phase adopts a Heusler-type structure, crystallizing in a cubic lattice within the $Fm\overline{3}m$ space group. In this configuration, the Fe atoms occupy two distinct coordination environments, denoted as Fe(1) and Fe(2). Each Fe(1) atom is bonded to four equivalent Fe(2) atoms and four equivalent Ge atoms, forming a distorted tetrahedral $FeFe_4Ge_4$ unit characterized by corner-, edge-, and face-sharing connectivity. The Fe–Fe and Fe–Ge bond distances within this unit are both 2.49 Å. The Fe(2) atoms occupy a 14-coordinate environment, bonded to eight equivalent Fe(1) atoms and six equivalent Ge atoms, with Fe–Ge bond lengths of 2.88 Å. Meanwhile, the Ge atoms are coordinated by six Fe(2) and eight Fe(1) atoms, resulting in a distorted body-centered cubic coordination geometry. An optical microscopy image of the $Fe_3Ge$ nanoplate grown on sapphire substrate is shown in Figure 1b, revealing a hexagonal flake morphology with a maximum lateral size of 78 μm. High-quality, large area nanoplates were successfully synthesized across a wide temperature range from 500°C to 750°C (see Figure S1). Figure 1e presents the atomic force microscopy (AFM) image of a representative $D0_3$-$Fe_3Ge$ nanoplate. The height profile illustrates a thickness of 102 nm and confirms the smooth surface morphology of the sample. Furthermore, by carefully optimizing the growth parameters, we achieved controlled synthesis of single crystalline nanoplates with thicknesses down to approximately 60 nm.

To verify the crystal structure of the CVD-grown-$Fe_3Ge$ nanoplates, we employed X-ray diffraction (XRD), scanning electron microscopy (SEM), and X-ray photoelectron spectroscopy (XPS). The XRD pattern (Figure 1f) shows three distinct diffraction peaks at 44.5°, 64.7°, and 81.8°, which are in excellent agreement with the standard reference for the $D0_3$ phase (PDF#04-004-6816). The SEM imaging further confirms a clean surface and high crystallinity (Figure S2). Figure 1g,h show the high-resolution XPS spectra of the $Fe_3Ge$ samples. Notably, due to the strong chemical activity of germanium and iron, a thin native oxide layer inevitably forms on the surface. While the XPS data indicates the presence of surface oxides, the signature of the underlying $Fe_3Ge$ phase remains prominent. The Fe2*p* spectrum (Figure 1g) can be deconvoluted into five distinct peak features: metallic Fe1 ($2p_{3/2}$), oxidized Fe2 ($2p_{3/2}$), a satellite peak, Fe3 ($2p_{1/2}$) peak, and a second satellite peak, located at 706.4 eV, 707.75 eV, 710.25 eV, 719.625 eV, and 722 eV, respectively. The Fe1 peak is sharp and symmetric, corresponding to the metallic $Fe^0$ state. The Fe2, Fe3 peaks are attributed to surface oxides, such as FeO and $Fe_2O_3$.[38] The Ge3*d* XPS spectrum (Figure 1h) provides further insights into the chemical environment. The spectrum was fitted into three primary peaks. The first peak, labeled Ge1 and located at 29.1 eV, exhibits a negative shift of 0.3 eV compared to elemental $Ge^0$ (29.4 eV). In $Fe_3Ge$, because the electronegativity of Ge (2.01) is higher than that of Fe (1.83), Ge acts as an electron acceptor. The resulting increase in electron density around the Ge atoms leads to a negative shift in binding energy, confirming its role as an electron acceptor with a formal valence state of about -4. The other two peaks, at 31.0 eV and 34.5 eV, correspond to the Ge ions in GeO and $GeO_2$, respectively.[39–41]

**Microstructure Characterization and Symmetry Analysis**

To further verify the composition and phase structure of the $Fe_3Ge$ nanoplates, we performed energy-dispersive X-ray spectroscopy (EDS), selected-area electron diffraction (SAED), and annular dark-field aberration-corrected scanning transmission electron microscopy (ADF-STEM). The $Fe_3Ge$ nanoplate sample grown on sapphire was prepared by using the focused ion beam (FIB) technology. As shown in Figure 2a, the atomic-resolution high-angle annular dark-field (HAADF)-STEM image along the [111] zone axis clearly reveals the characteristic hexagonal honeycomb lattice of the $D0_3$ structure. The measured interplanar spacings of 2.04 Å correspond to the (2, -2 ,0) and (2, 0, -2) planes of $Fe_3Ge$, demonstrating excellent agreement with the simulated image in Figure 2g. Furthermore, the atomic configuration is well corroborated by the structural model (Figure 2b). In this projection, the bright spots correspond to atomic columns composed of both Fe and Ge; these columns exhibit nearly identical intensities and arrangements in both experimental and simulated images. EDS mapping confirms a uniform distribution of Fe and Ge throughout the nanoplate (Figure 2c,d). The lattice constants extracted from the HAADF-STEM data of the region of the nanoplate (top area in Figure 3a) are $a = b = c \approx 5.77$ Å, closely matching the ideal lattice parameter (5.75 Å) for $D0_3$-$Fe_3Ge$. This suggests high-quality epitaxial growth on the sapphire substrate. The high crystalline quality of the CVD-grown $Fe_3Ge$ nanoplate is further evidenced by the homogeneous atomic arrangement and the absence of localized defects in the STEM images. Furthermore, the experimental SAED pattern (Figure 2f) and the corresponding simulation (Figure 2h) exhibit an identical set of hexagonally arranged diffraction spots. The two brightest spots can be indexed to the (2, -2, 0) and (2, 0, -2) planes, respectively. This high symmetry is fully consistent with the $Fm\bar{3}m$ space group of $Fe_3Ge$, confirming the single-crystalline nature of the nanoplate with a hexagonal morphology. However, cross-sectional STEM analysis reveals a distinct structural evolution along the growth direction (Figure 3a). The explicit crystallographic epitaxial relationship between the $D0_3$-$Fe_3Ge$ nanoplate and the c-plane sapphire substrate can be defined as (111) plane of $D0_3$-$Fe_3Ge$ parallel to (0001) plane of c-plane sapphire for the out-of-plane orientation. At the $Fe_3Ge$/sapphire interface, the interplanar spacing is compressed to 1.955 Å (Figure 3c), corresponding to a significant lattice mismatch. This implies that the initial growth layers are subjected to a giant compressive strain of ~4.16% to accommodate the substrate. As the thickness increases, this strain gradually relaxes, restoring the lattice constant to the value of 2.04 Å near the surface (Figure 3b). This observed strain gradient confirms an epitaxial growth mode where interfacial stress is released through lattice relaxation.

Second-harmonic generation (SHG) measurements were further employed to probe the crystallographic and magnetic symmetries due to the high sensitivity of this method to the crystal symmetry and orientation. The observation of rotational anisotropy in the SHG intensity (Figure 3d) provides profound insights into the crystal symmetry of the CVD-grown nanoplates. Although various factors can potentially contribute to the SHG signal in nanoplates (e.g., sapphire substrate, surface electric-dipoles, electric-quadrupoles, and magnetic dipoles), we have decoupled these origins. First, the measurement on the bare sapphire substrate yielded a negligible SHG intensity, confirming that the observed SHG signal originates from the $D0_3$-$Fe_3Ge$ nanoplates. Regarding the surface electric-dipole contribution, it typically originates only from the topmost few atomic layers of the film, whereas the strain-induced bulk electric-dipole response is generally much stronger than this surface signal. Furthermore, in materials with high dielectric constant such as metallic, electric-quadrupole act as higher-order perturbations, and its SHG intensity is weaker than that of the surface electric-dipole.[42] Finally, regarding the magnetic-dipole contribution, our magnetic measurement in the following text reveals that $D0_3$-$Fe_3Ge$ exhibits soft magnetic properties. Since our optical measurements were performed under zero magnetic field, the magnetic domains within the film were randomly distributed. This random orientation leads to no net macroscopic magnetization, which results in no magnetic-dipole contribution. Since bulk $D0_3$-$Fe_3Ge$ belongs to the centrosymmetric $Fm\bar{3}m$ space group, electric-dipole SHG is strictly forbidden. Therefore, the SHG results provide possible evidence for strain-induced symmetry breaking. Although STEM shows that the strain gradually relaxes along the thickness direction, this strong epitaxial strain may have a profound impact on the overall magnetic transport properties.

**Magnetic Properties and Anisotropy**

To investigate the magnetic properties of the CVD-grown $D0_3$-$Fe_3Ge$ nanoplates, we conducted measurements using a superconducting quantum interference device (SQUID) magnetometer. Figure 4a,b show the magnetization hysteresis (*M*-*H*) loops measured in-plane and out-of-plane at temperatures ranging from 4 K to 300 K. The loops exhibit clear ferromagnetic characteristics with well-defined saturation, confirming the robust ferromagnetic order in our samples. To extract the true saturation magnetization ($M_s$) from the high-field data, we fitted the magnetization ($M$) using the law of approach-to-saturation:

$$M = M_s\left(1 - \frac{a}{H} - \frac{b}{H^2}\right) + \chi_p H \quad (1)$$

In this model, $M_s$ represents the absolute saturation magnetization. The term $a/H$ is primarily attributed to magnetic anisotropy, while the $b/H^2$ term originates from microstructural contributions such as stresses and crystal defects. The final term, $\chi_p H$, accounts for the field-induced Pauli paramagnetic susceptibility of the conduction electrons. The saturation magnetization at 4 K was determined to be approximately $8.012\times10^{-3}$ emu in the in-plane direction and $5.615\times10^{-3}$ emu in the out-of-plane direction. A more striking contrast is observed at a low field of 0.5 kOe, where the out-of-plane moment is only $0.17\times10^{-3}$ emu (3.1% of its $M_s$), whereas the in-plane moment reaches $4.13\times10^{-3}$ emu (51.5% of its $M_s$). This disparity indicates clear magnetic anisotropy. Furthermore, as shown in Figure 4c, the saturation field in the out-of-plane direction is significantly larger than that in the in-plane direction at 300 K, indicating that the magnetic easy-axis lies in-plane (perpendicular to the *c*-axis). This indicates that the magnetic moments of the D0$_3$-$Fe_3Ge$ nanoplates tend to align in-plane to minimize the total free energy of the system. In its bulk form, D0$_3$-$Fe_3Ge$ belongs to the cubic crystal system and exhibits a relatively weak intrinsic magneto crystalline anisotropy. Since shape anisotropy and demagnetization effects naturally favor in-plane magnetization for thin film,[43] we quantitatively evaluated the demagnetization effect induced by the film geometry. Since the result is merely a semi-quantitative estimation, it is only discussed in terms of relative trends. Assuming an average nanoplate thickness (d) of 150 nm and a coverage (η) of 50%, the extracted volume saturation magnetization $M_{s,V} = M_s/V = M_s/(\mathrm{S} \times \mathrm{d} \times \eta) \approx$ 1068 emu/cm$^3$,which is corresponding to ~ 5.53 $\mu_B$/f.u. and approximately consistent with bulk (~ 5.29 $\mu_B$/f.u.).[35] Given the extremely large aspect ratio of the nanoplates, the in-plane demagnetization factor approaches zero, while the out-of-plane demagnetization factor approaches its maximum 4π. This yields a shape anisotropy energy $K_s = 2\pi M_{s,V}^2 \approx 7.17 \times 10^6$ erg/cm$^3$ ( ~ 0.22 meV/f.u.), establishing a substantial energy barrier and indicating an exceptionally strong demagnetization effect. Subsequently, we extracted the effective anisotropy energy $K_{eff}$ from the *M-H* curves using the following equation:

$$K_{\mathrm{eff}} = \frac{1}{V}\left[\int_0^{M_s} (H_\perp - H_\parallel) dM\right]$$

The calculation yields $K_{eff} \approx 5.11 \times 10^6\ \mathrm{erg/cm^3}$ at 4 K and $K_{eff} \approx 3.68 \times 10^6\ \mathrm{erg/cm^3}$ at 300 K. These values are of the same order of magnitude and closely match. Furthermore, we calculated the magnetocrystalline anisotropy energy $E_{MAE}$ for bulk $D0_3$-$Fe_3Ge$ with and without strain. The calculated $E_{MAE} = (E_\perp - E_\parallel)$ is $-0.05$ meV/f.u. without strain and shifts to $-0.016$ meV/f.u. under a 4.16% strain. This intrinsic $E_{MAE}$ is far smaller than the demagnetization energy, demonstrating that the demagnetization effect is the primary reason of the magnetic anisotropy.

The high Curie temperature ($T_c$) of these $D0_3$-$Fe_3Ge$ nanoplates, which is well above room temperature, is inferred from the gradual decay of the saturation moment shown in Figure 4d. This thermal stability is consistent with previous reports and confirms the material's potential for ambient-temperature spintronic applications[34,44–46]. The remanence ratio ($M_r/M_s$), where $M_r$ is the remanent magnetization, is notably low, measured at approximately 0.0023 in the in-plane direction at 300 K, suggesting that the macroscopic magnetization does not remain fully aligned once the external field is removed. The relatively low remanence ratio, which is advantageous for spintronic applications requiring efficient switching, can be attributed to the material's magnetic softness and thin-film geometry. These factors promote domain formation at zero field to minimize magnetostatic energy. At room temperature, the measured coercivity ($H_c$) is as low as 10 Oe (see Figure S11), a quintessential characteristic of soft magnets, that indicates very small external fields are sufficient to reverse the magnetization. This soft magnetic behavior reflects the high quality of our samples and a low density of defects.

The reduction in magnetic flux pinning centers not only facilitates easy magnetization reversal but also suppresses extrinsic electron scattering. Consequently, the intrinsic Berry curvature-driven AHE emerges as the dominant mechanism, as supported by the scaling law analysis in the following section. Such properties are highly desirable for applications in magnetic sensors and memory devices, as they directly translate to lower operational power consumption.

**AHE**

To further analyze the ferromagnetic character of $Fe_3Ge$ nanoplates, the Hall devices on as-grown $Fe_3Ge$ nanoplates on sapphire substrates were fabricated to investigate the AHE through the magneto-transport measurements. We focus on the longitudinal ($R_{xx}$) and transverse ($R_{xy}$) resistance as well as their ratio to analyze the intrinsic physical mechanisms and magnetic properties. This approach avoids uncertainties associated with geometric factors in unpatterned nanoplates while preserving the pristine quality of the CVD-grown interface. Figure 5a presents a schematic diagram of the $Fe_3Ge$ device on sapphire substrate. The magnetic field was applied perpendicularly to the sample surface, while the current flowed in-plane. The temperature-dependent longitudinal resistance for $Fe_3Ge$ nanoplates was measured from 300 K to 1.6 K in a zero magnetic field during the cooling process are shown in Figure 5b. The $R_{xx}$ increases monotonically within the temperature range of 30 K to 300 K, demonstrating the metallic properties. However, a resistance upturn emerges below 30 K. This phenomenon can be elucidated by the weak localization theory. At high-temperatures, the electron phase coherence length is shortened by inelastic phonon scattering, rendering electron interference effects negligible and thus leading to metallic conduction. Conversely, in the low-temperature regime, impurity-induced elastic scattering coupled with the weak localization effect causes the film to display non-metallic properties.[37,47] Figure 5c shows $R_{xy}$ measured at temperatures from 1.6 K to 300 K. The $R_{xy}$ shows a sharp change followed by the saturation, a hallmark of ferromagnetism in $Fe_3Ge$. At magnetic fields over the saturation field ($H_{sat}$), $R_{xy}$ exhibits a linear dependence on magnetic fields, which is attributed to the ordinary Hall effect (OHE). The observed positive slope indicates that the majority charge carriers are holes. The hole carrier concentration ($n$) is estimated to be approximately $10^{23}$ $cm^{-3}$. This high carrier density also corroborates the typical metallic nature of the $D0_3$-$Fe_3Ge$ phase. Furthermore, the saturation field shows negligible variation with temperature, signifying the stability of the robust AHE of the $Fe_3Ge$ nanoplates, which persists far beyond room temperature. The Hall resistance ($R_{AHE}$) can be determined by extrapolating the saturation state back to zero magnetic field. Figure 5e presents the scaling behavior of $R_{AHE}$ against $R_{xx}$ in a log-log scale. A linear fitting was performed across the entire measured temperature range. According to the theory of AHE, a scaling exponent of $\alpha \approx 2$ strictly rules out the extrinsic skew-scattering mechanism (which requires $\alpha = 1$).[48] Instead, it points towards the intrinsic mechanism (linked to the Berry curvature) or the side-jump mechanism as the dominant physical origin. To further distinguish between the intrinsic and side-jump mechanisms, we investigated the resistance ratio $R_{AHE}/R_{xx}^2$, which is proportional to the anomalous Hall conductivity $\sigma_{\mathrm{AHE}}$, as a function of temperature in Figure 5f. For $\sigma_{\mathrm{AHE}} \propto g \cdot R_{\mathrm{AHE}}/R_{\mathrm{xx}}^2$, where $g$ represents a geometric factor that does not change with temperature, the temperature dependence of AHC can be accurately captured by the resistance ratio. $R_{AHE}/R_{xx}^2$ exhibits a remarkably weak temperature dependence over a wide temperature range in Figure 5f. The robust behavior of AHC with respect to temperature indicates that the Hall effect is not dominated by the scattering processes. In Figure S13, AHC shows negligible changes with $\sigma_{xx}$. This behavior strongly suggests that the intrinsic mechanism, originating from the integration of the Berry curvature $\Omega(k)$ over the occupied Brillouin zone, is the dominant contributor to the AHE.[49,50] The intrinsic AHC $\sigma_{\mathrm{AHE}}^{\mathrm{int}}$ can be expressed as:

$$\sigma_{\mathrm{AHE}}^{\mathrm{int}} = -\frac{e^2}{\hbar}\sum_n \int_{BZ} \frac{dk}{(2\pi)^d} f_n(k)\Omega_{n,z}(k) \quad (2)$$

where $f_n(k)$ is the Fermi-Dirac distribution function for the n-th band. Crucially, the Berry curvature $\Omega_{n,z}(k)$ is an intrinsic property of the band structure, largely independent of scattering processes.[48,51] The Berry curvature itself is a geometric property of the wavefunction, which is determined by the electronic structure at T = 0 K. It does not exhibit significant variation with temperature at scales below the characteristic electronic energy (e.g., band gap). The primary effect of increasing temperature is the thermal broadening of the Fermi distribution function $f_n(k)$. However, provided the Berry curvature is slowly varying around the Fermi energy and sharp topological features (such as avoided crossings or Weyl nodes) reside well away from the Fermi level, the intrinsic anomalous Hall contribution is robust against thermal smearing of the Fermi-Dirac distribution. In typical ferromagnetic materials, the dominant contribution of AHE either stems from the skew scattering of extremely pure samples or from the strong intrinsic Berry curvature contribution. The magnitude of the side-jump is only $(e^2/ha)(\varepsilon_{\mathrm{SO}}/E_{\mathrm{F}})$, which can be considered as a background perturbation. $\varepsilon_{\mathrm{SO}}$, $E_{\mathrm{F}}$, $e$, $h$ and $a$ represents spin-orbit interaction energy, fermi energy, electronic charge, Planck constant, and lattice parameter, respectively.[3,30] For ferromagnetic metals, $\varepsilon_{\mathrm{SO}}/E_{\mathrm{F}}$ is generally less than $10^{-2}$.[52] After calculation, it can be estimated that the contribution of the side jump is approximately 10 to 30 $\Omega^{-1}\cdot\mathrm{cm}^{-1}$, which is far less than the contribution of the intrinsic mechanism.

**Theoretical Calculation**

To reveal the differences between our as-grown $D0_3$-$Fe_3Ge$ nanoplates and their bulk counterparts, we performed density functional theory (DFT) calculations to investigate the effect of strain on the transport properties of the $D0_3$-$Fe_3Ge$ nanoplates. As shown in Figure 6a, under strain-free conditions, $D0_3$-$Fe_3Ge$ exhibits an anomalous Hall conductivity (AHC) of approximately 288 $\Omega^{-1}\cdot\mathrm{cm}^{-1}$, which is consistent with previous reports.[35] Along the high symmetry paths Γ-X and W-X, although band crossings and spin-orbit coupling (SOC) induced gaps exist, these prominent crossing points are located far from the Fermi level. Instead, the Fermi level traverses relatively flat bands that contribute minimally to the Berry curvature.

In contrast, the epitaxially grown nanoplates experience a 4.16% interfacial compressive strain. This strain leads to a reduced lattice constant and shortened Fe-Fe distances, which consequently enhances the exchange interaction. Simultaneously, an upward shift of the bands along the W-X path is clearly observed, accompanied by the reconstruction of certain bands near the L-W path. This evolution results in the emergence of more band crossing points at the Fermi level and further leads to an enhanced density of states near the Fermi surface. Crucially, under 4.16% compressive strain, the SOC-induced anti-crossing gaps along the X-U and W-X paths shift to within ±0.1 eV of the Fermi level, with gap sizes of 15–40 meV. This proximity to $E_F$, in contrast to the unstrained case where such gaps lie more than 0.3 eV away, enables the Berry curvature to contribute maximally to the transverse transport. The calculated AHC for this strained state reaches up to 550 $\Omega^{-1}\cdot cm^{-1}$. Disregarding the errors introduced by the irregular device geometry, our Hall devices yielded measured AHC values of 571.3 $\Omega^{-1}\cdot cm^{-1}$, 573.7 $\Omega^{-1}\cdot cm^{-1}$, and 592.8 $\Omega^{1}\cdot cm^{-1}$ at 1.6 K, 100 K, and 300 K, respectively. Although the DFT calculation assumes a uniform strain structure, which may differ from the actual situation of gradual relaxation. We calculated the AHC of $D0_3$-$Fe_3Ge$ from no strain to 5% strain (see Figure S22). This leap in AHC demonstrates that epitaxial strain serves as a powerful tuning mechanism, effectively unleashing the contribution of the Berry curvature in momentum space to transport properties.

# 3. Conclusion

In summary, we successfully synthesized non-layered $D0_3$-$Fe_3Ge$ nanoplates with room-temperature ferromagnetism by CVD method. Structural analysis suggested by XRD, XPS, AFM and STEM measurements confirmed the high crystalline quality of the $D0_3$-$Fe_3Ge$ nanoplates. Unlike their polycrystalline bulk counterparts, these nanoplates exhibit a unique structural state characterized by 4.16% epitaxial strain. Crucially, this interfacial strain might break the centro-symmetry of the cubic lattice, as SHG showed evident nonlinear optical signals. Magnetic measurements using SQUID revealed clear magnetic anisotropy and an in-plane easy-axis. The low coercivity and low remanence ratio indicate facile flipping of magnetic moments, while the high saturation magnetization maintained at room temperature confirms ferromagnetic ordering with a Curie temperature well above 300 K, underscoring its potential for room-temperature applications. Furthermore, magneto-transport measurements on Hall devices demonstrated the robust AHE and strongly suggest the intrinsic origin of the AHE in CVD-grown $Fe_3Ge$. This study not only advances the synthesis of magnetic materials but also establishes $D0_3$-$Fe_3Ge$ nanoplates as a compelling platform for exploring dimensionally confined magnetism. The robust intrinsic AHE at room temperature suggests promising prospects for ultra-low power spintronic devices.

## 4. Experimental Section

**CVD Synthesis of Ferromagnetic $D0_3$-$Fe_3Ge$ nanoplates**

$FeCl_2$ (8-10 mg, Alfa, 99.5%) and $GeI_2$ (3-5 mg, Aladdin, 99.99%) mixed powder were used as precursors placed on the middle of quartz boat. The sapphire substrate was placed directly above the powder. Before the heating process, high purity argon (Ar, 500 sccm) gas was used to purge the quartz tube for ten minutes. Subsequently, the furnace was ramped to the target temperature (500 ℃ to 750 ℃) at a rate of 100 ℃ per minute. 25 sccm of hydrogen($H_2$) and 100 sccm Ar were used as the carrier gas. The growth process lasted for 3 to 5 minutes. After growth, the furnace was naturally cooled to the room temperature.

**Structure and Composition Characterization**

The crystalline structure feature of $Fe_3Ge$ was characterized by an X-ray diffraction system (XRD) (Rigaku D/MAX-Ultima III, Japan, Cu-Kα radiation $\lambda = 1.5406$ Å). The chemical composition and elemental valence of the crystal were characterized by an X-ray photoelectron spectroscopy (PHI5000 Versa Probe, Japan). SHG measurements were conducted with 810 nm laser excitation. An incident power of 3.6 mW was used for measurements. The surface morphology and element distribution of $Fe_3Ge$ were characterized by scanning electron microscopy (SEM) (Carl Zeiss, Gemini SEM 500). The atomic structure of $Fe_3Ge$ was characterized by high angle annular dark-field scanning transmission electron microscopy (ADF-STEM) imaging using a probe aberration-corrected STEM (Titan cubed G2 60-300) at 300 kV. The $Fe_3Ge$ sample for STEM measurements was prepared by FIB (FEI Helios G4 CX).

**Magnetic Properties Measurement**

SQUID-MPMS measurements were conducted using the Quantum Design magnetic property measurement system (MPMS(SQUID)XL-7). *M-H* magnetic tests were conducted on the $Fe_3Ge$ samples deposited on the sapphire substrates. The magnetic field ranging from -3 T to 3 T was applied in both the in-plane and out-of-plane directions. The temperature range was from 4K to 300K.

**Hall Device Fabrication and Measurements**

The devices based on $Fe_3Ge$ were fabricated on the sapphire substrates by e-beam lithography, then Ti/Au (5/60 nm) contacts were deposited by e-beam evaporation. The devices were measured in an Oxford Instruments magnet system, with the magnetic field applied perpendicular to the plane of the $Fe_3Ge$ nanoplate. The direction of the applied current was parallel to the edge of the hexagonal nanoplate. The crystal orientation of edge corresponds to the <110> group (i.e. $[1\bar{1}0]$, $[01\bar{1}]$ and $[10\bar{1}]$).

**Theoretical Calculation**

First-principles calculations of the band structure for $D0_3$-$Fe_3Ge$ nanoplates were performed within the framework of density functional theory (DFT),[53] as implemented in the Vienna Ab initio Simulation Package (VASP).[54,55] The projector-augmented-wave (PAW) pseudopotentials[56] were utilized to model the electron-ion interactions, and the Perdew-Burke-Ernzerhof (PBE) parametrized generalized gradient approximation (GGA) was employed to describe the exchange-correlation potentials.[57] The AHC was subsequently evaluated using the Wannier90[58] and WannierTools package.[59] The tight-binding models were built in WANNIER90 through maximally localized Wannier functions by considering the *d* orbitals of Fe and *p* orbitals of Ge. The kinetic energy cutoff and the charge density cutoff were set to 500 eV. The charge densities were determined self-consistently on an unshifted k-mesh of 12×12×12 for geometric structures calculations of $D0_3$-$Fe_3Ge$. The convergence criteria were $1\times10^{-8}$ eV for the energy difference in electronic self-consistent calculation and $1\times10^{-2}$ eV/Å for the residual forces on atoms. Strain is defined as (a-a0)/a0×100%, where a0 and a represent the lattice constants. $D0_3$-$Fe_3Ge$ has ferromagnetic ground state and the direction of the magnetic moment used for calculation is perpendicular to the plane direction. The SOC effect was incorporated through the non-collinear magnetic calculation module implemented in the VASP, utilizing fully relativistic pseudopotentials. SOC was self-consistently included in the Hamiltonian throughout both the band-structure calculations and the integration of the AHC. The tight-binding models, constructed via maximally localized Wannier functions as implemented in the Wannier90 package, fully retained the Berry curvature corrections, thereby ensuring the reliability of the AHC evaluations.

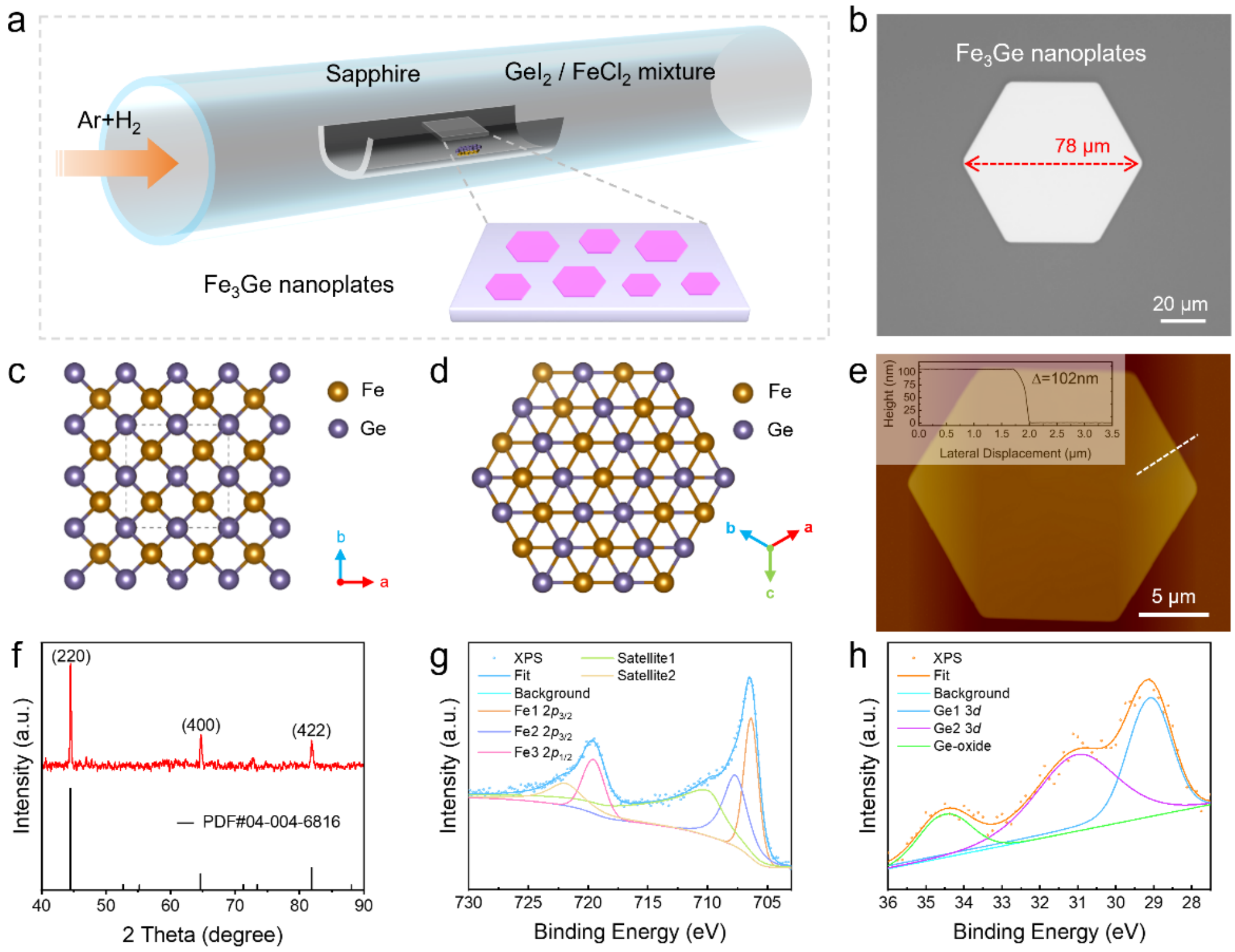

**Figure 1.** Synthesis and characterization of D0$_3$-Fe$_3$Ge. (a) Schematic diagram of the CVD growth approach for the synthesis of Fe$_3$Ge nanoplates. (b) An optical microscopy image of Fe$_3$Ge grown on a sapphire substrate. (c,d) Atomic structure of D0$_3$-Fe$_3$Ge crystal. (e) Atomic force microscopy image of D0$_3$-Fe$_3$Ge nanoplate. (f) XRD pattern of as-grown Fe$_3$Ge nanoplates. (g,h) High-resolution XPS spectra of Fe2*p* (g) and Ge3*d* (h) of Fe$_3$Ge.

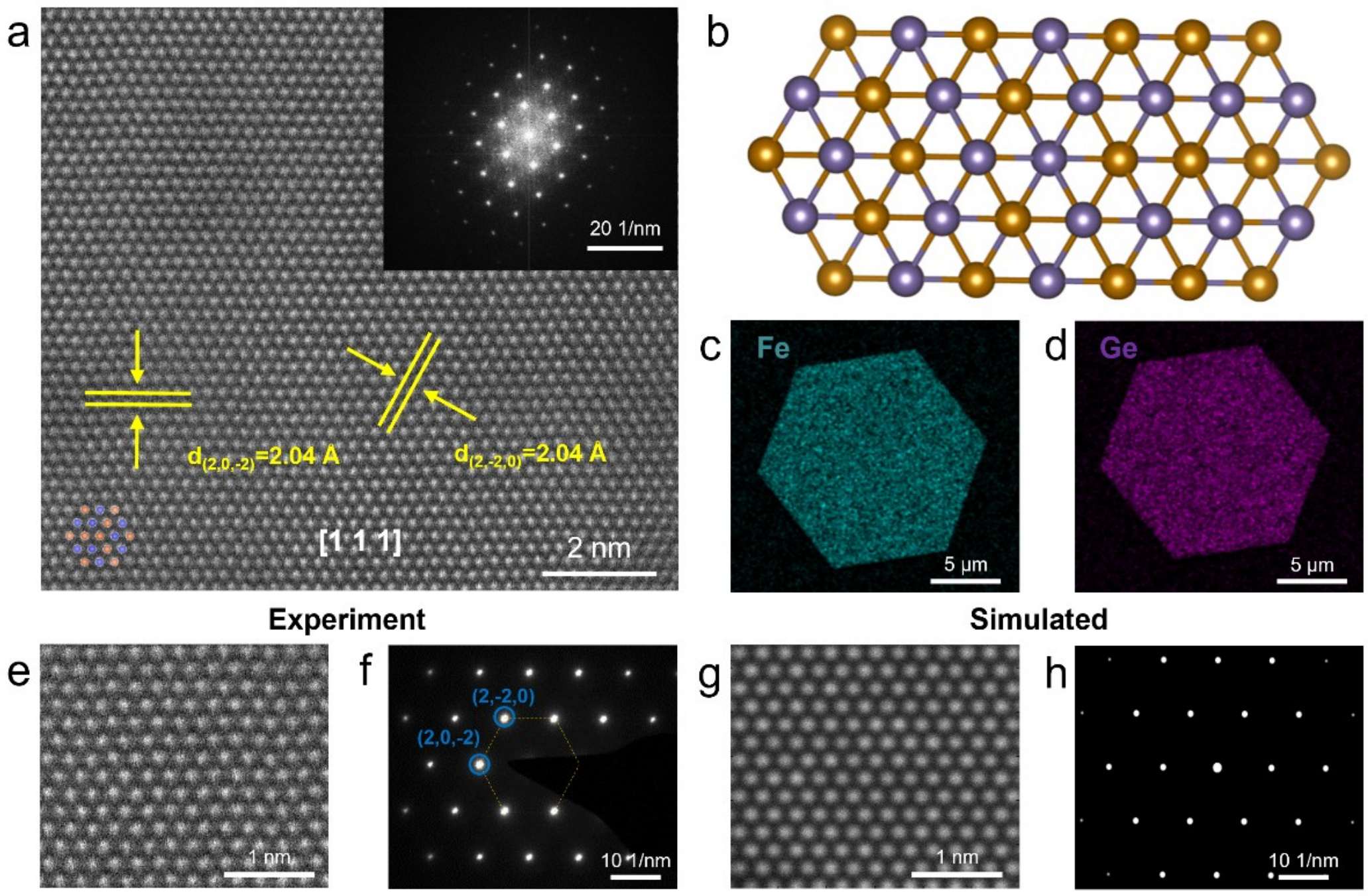

**Figure 2.** Composition and microstructure characterization of CVD-grown $Fe_3Ge$ nanoplates. (a) High resolution of STEM-HAADF image of $Fe_3Ge$ along the [1 1 1] zone axis. Both the lattice constants of the (2,0,-2) and (2,-2,0) are 2.04 Å. The inset in the upper right corner is the corresponding FFT. (b) Atomic structure diagram of $Fe_3Ge$ along the [1 1 1] zone axis. (c,d) EDS mapping of Fe (c) and Ge (d). (e,f) Experimentally measured STEM-HAADF image (e) and SAED pattern (f) of the $Fe_3Ge$ crystal. (g,h) Simulated STEM-HAADF image (g) and SAED pattern (h) of the $Fe_3Ge$ crystal.

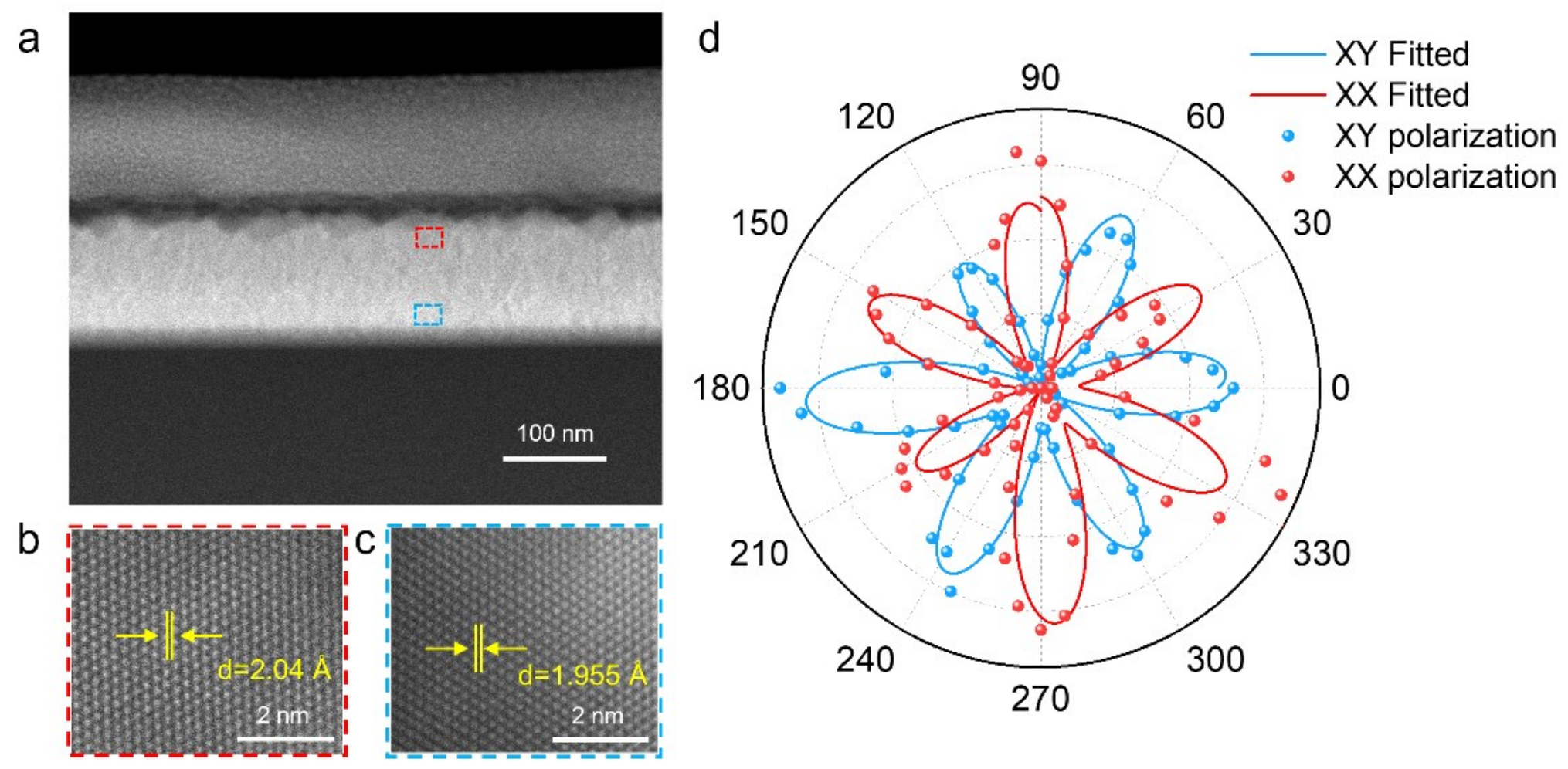

**Figure 3.** Comparison of the interplanar spacings measured from STEM images taken at different regions (top vs. bottom) of the nanoplate. (a) Low-magnification cross-sectional HAADF-STEM image of D0$_3$-$Fe_3Ge$ nanoplate. (b,c) Interplanar spacings of top and bottom regions. (d) Polarization angle-dependent SHG intensity of $Fe_3Ge$ nanoplates under parallel (XX) and perpendicular (XY) polarization at 10 K.

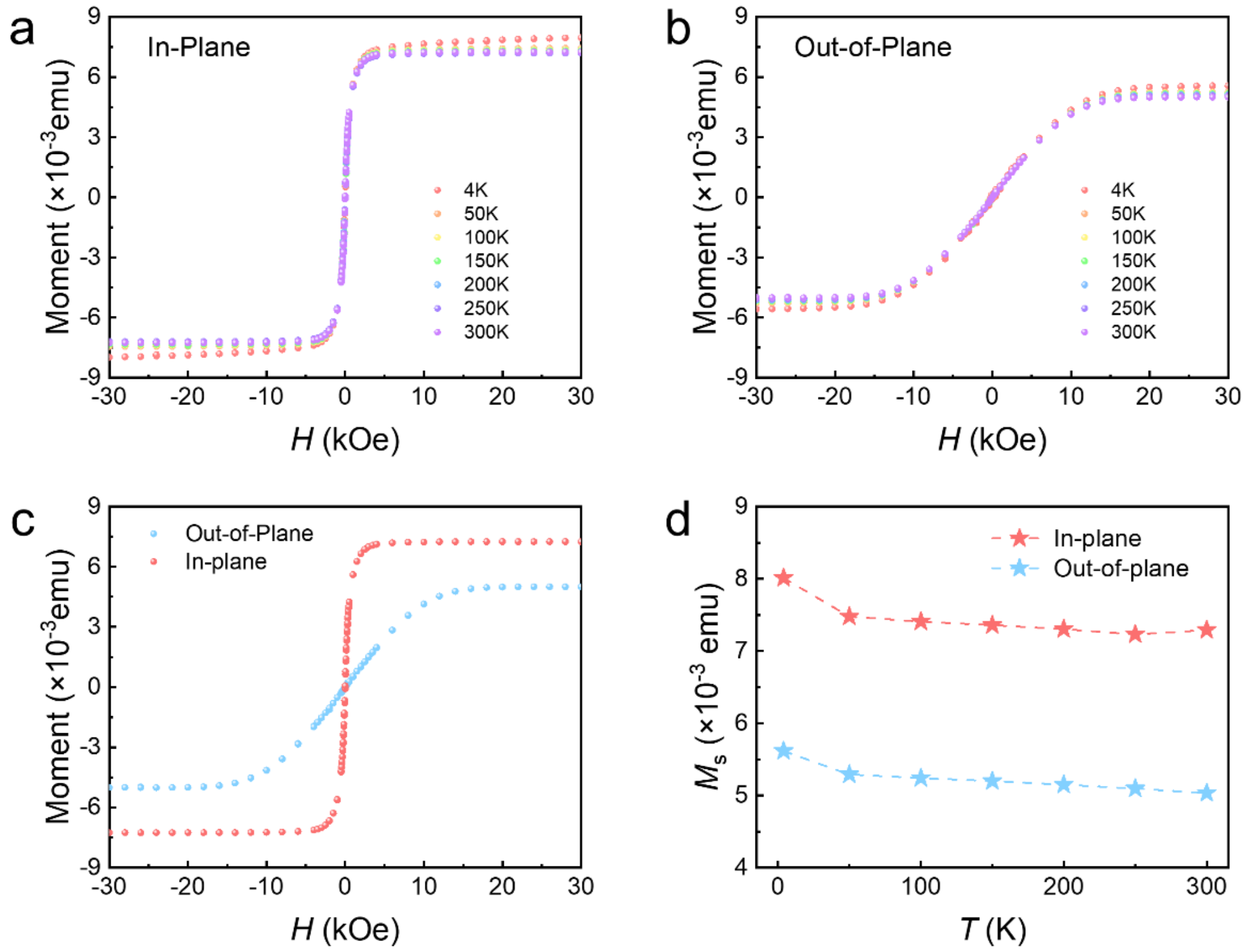

**Figure 4.** Magnetic measurement of D0$_3$-$Fe_3Ge$. (a,b) Field-dependent magnetization of $Fe_3Ge$ nanoplates measured in-plane (a) and out-of-plane (b) respectively, at specific temperature ranging from 4 K to 300 K. (c) Comparison of magnetization of $Fe_3Ge$ nanoplates measured in-plane and out-of-plane under 300K. (d) Temperature dependence of the saturation magnetization of $Fe_3Ge$ nanoplates.

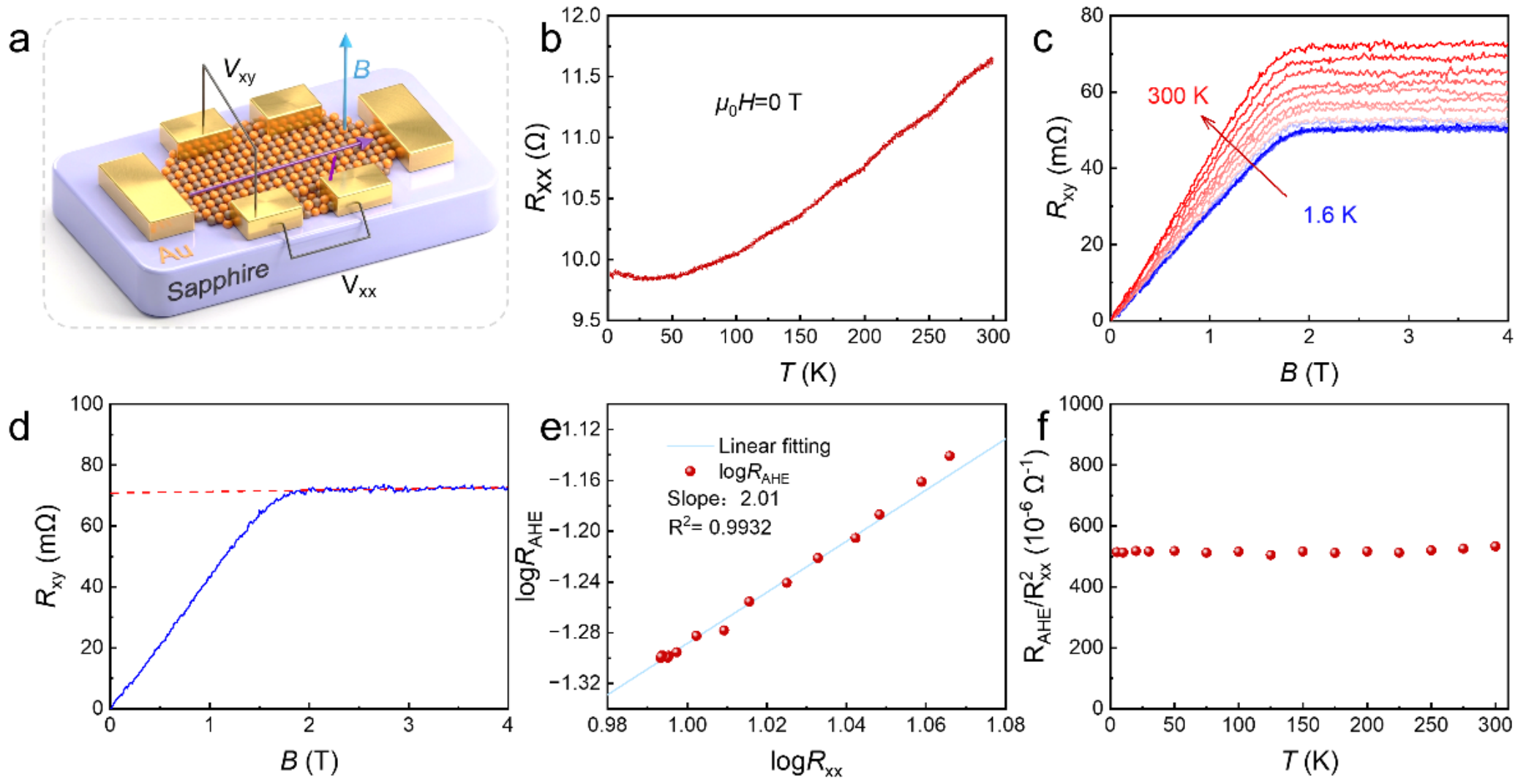

**Figure 5.** AHE of D0$_3$-Fe$_3$Ge. (a) The schematic diagram of the $Fe_3Ge$ Hall device fabricated on a sapphire substrate. $I$, $B$, $V_{xx}$ and $V_{xy}$ represent drive current, magnetic field, longitudinal voltage, transverse Hall voltage respectively. (b) Temperature dependence of the $R_{xx}$ measured under zero magnetic field. (c) Temperature dependence of the $R_{xy}$ of $Fe_3Ge$ nanoplates ranging from 1.6 K to 300 K. (d) Extrapolate the Hall resistance to zero magnetic field to derive the anomalous Hall resistance under the current of 4 μA at 300K. The slope of the high field linear regions is 0.467 mΩ/T. (e) The variation of log$R_{AHE}$ with log$R_{xx}$ and its linear fitting for $Fe_3Ge$. The slope of linear fitting is 2.01 (f) Temperature dependence of $R_{AHE}/R_{xx}^2$.

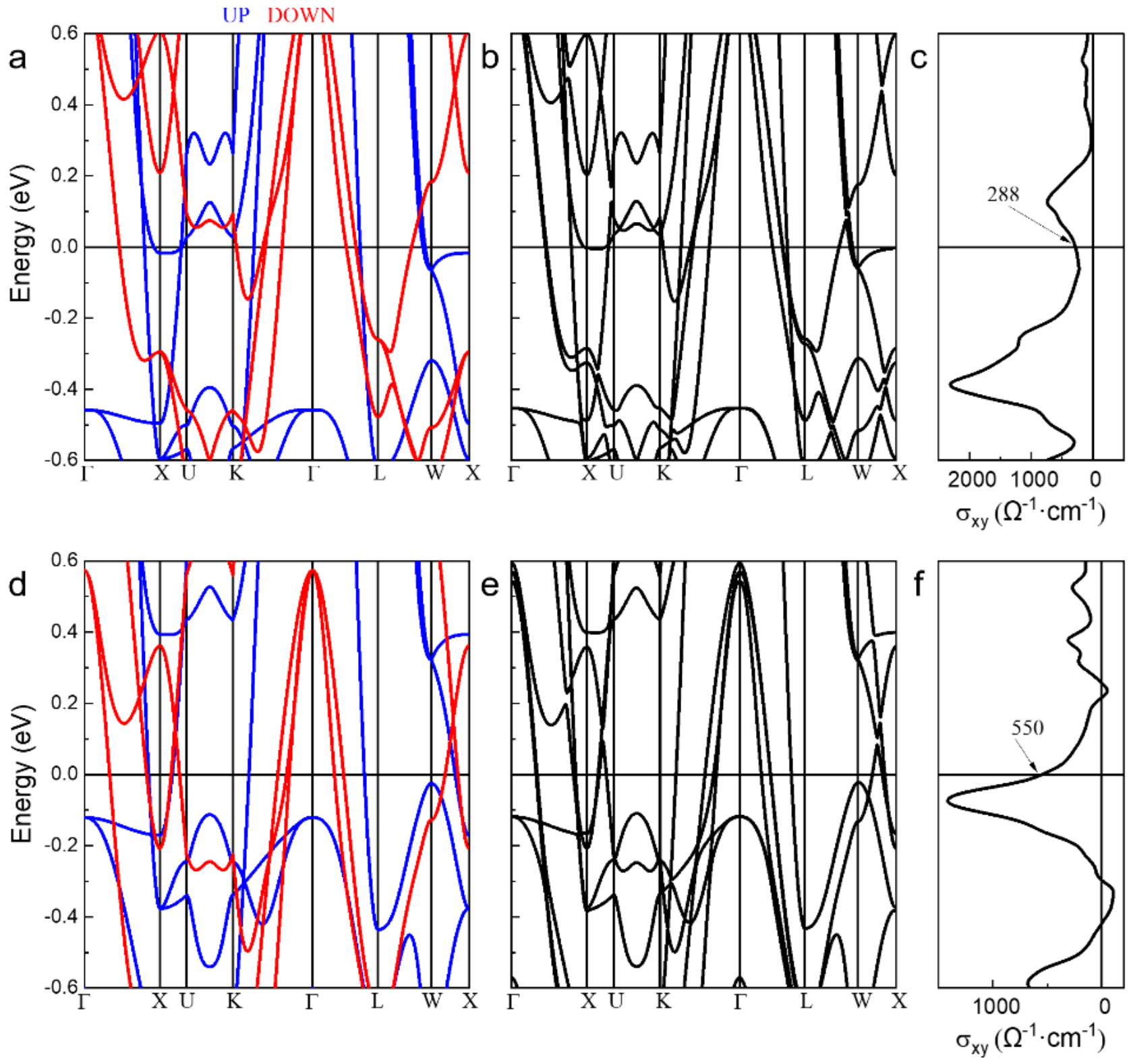

**Figure 6.** Band structure without SOC (a), with SOC (b) and calculated AHC (c) for unstrained $D0_3$-$Fe_3Ge$. Band structure without SOC (d), with SOC (e) and calculated AHC (f) for $D0_3$-$Fe_3Ge$ with 4.16% strain.

# ASSOCIATED CONTENT

**Data Availability Statement**

The data analyzed during the current study available from the corresponding author on reasonable request.

**Supporting Information**

The Supporting Information is available: Additional characterizations regarding $D0_3$-$Fe_3Ge$ and the specific details of the process of Hall signal.

# AUTHOR INFORMATION

**Corresponding Author**

Geliang Yu − National Laboratory of Solid State Microstructures, Collaborative Innovation Center of Advanced Microstructures, School of Physics, Nanjing University, Nanjing 210093, China;

International Joint Laboratory on Two Dimensional Materials, Nanjing University, Nanjing 210093, China;

Jiangsu Physical Science Research Center, Nanjing University, Nanjing 210000, China;

Email: yugeliang@nju.edu.cn

Lei Wang − National Laboratory of Solid State Microstructures, Collaborative Innovation Center of Advanced Microstructures, School of Physics, Nanjing University, Nanjing 210093, China;

International Joint Laboratory on Two Dimensional Materials, Nanjing University, Nanjing 210093, China;

Jiangsu Physical Science Research Center, Nanjing University, Nanjing 210000, China;

Email: leiwang@nju.edu.cn

Renjun Du − National Laboratory of Solid State Microstructures, Collaborative Innovation Center of Advanced Microstructures, School of Physics, Nanjing University, Nanjing 210093, China;

Email: renjundu@nju.edu.cn

**Authors**

Mingxuan Chen – National Laboratory of Solid State Microstructures, Collaborative Innovation Center of Advanced Microstructures, School of Physics, Nanjing University, Nanjing 210093, China

Chao Chen – National Laboratory of Solid State Microstructures, Collaborative Innovation Center of Advanced Microstructures, School of Physics, Nanjing University, Nanjing 210093, China

Siqi Jiang – National Laboratory of Solid State Microstructures, Collaborative Innovation Center of Advanced Microstructures, School of Physics, Nanjing University, Nanjing 210093, China

Shuxiang Qiao – School of Physics and Physical Engineering, Qufu Normal University, Qufu 273165, China

Hongyan Lu – School of Physics and Physical Engineering, Qufu Normal University, Qufu 273165, China

Di Zhang – National Laboratory of Solid State Microstructures, Collaborative Innovation Center of Advanced Microstructures, School of Physics, Nanjing University, Nanjing 210093, China

Jiabei Huang – National Laboratory of Solid State Microstructures, Collaborative Innovation Center of Advanced Microstructures, School of Physics, Nanjing University, Nanjing 210093, China

Fuzhuo Lian – National Laboratory of Solid State Microstructures, Collaborative Innovation Center of Advanced Microstructures, School of Physics, Nanjing University, Nanjing 210093, China

Zhaochen Qu – National Laboratory of Solid State Microstructures, Collaborative Innovation Center of Advanced Microstructures, School of Physics, Nanjing University, Nanjing 210093, China

Siqin Wang – National Laboratory of Solid State Microstructures, Collaborative Innovation Center of Advanced Microstructures, School of Physics, Nanjing University, Nanjing 210093, China

Shuai Zhang – National Laboratory of Solid State Microstructures, Collaborative Innovation Center of Advanced Microstructures, School of Physics, Nanjing University, Nanjing 210093, China

Yulu Liu – National Laboratory of Solid State Microstructures, Collaborative Innovation Center of Advanced Microstructures, School of Physics, Nanjing University, Nanjing 210093, China

Ronghua Liu – National Laboratory of Solid State Microstructures, Collaborative Innovation Center of Advanced Microstructures, School of Physics, Nanjing University, Nanjing 210093, China;

Jiangsu Provincial Key Laboratory for Nanotechnology, Nanjing University, Nanjing 210093, China;

National Key Laboratory of Spintronics, Nanjing University, Suzhou 215000, China;

Xiaoxiang Xi – National Laboratory of Solid State Microstructures, Collaborative Innovation Center of Advanced Microstructures, School of Physics, Nanjing University, Nanjing 210093, China;

**Author Contributions**

†M.C., C.C, S.J. and S.Q. contributed equally.

**Notes**

The authors declare no competing financial interest.

# ACKNOWLEDGMENTS

The authors would like to thank the International Joint Lab of 2D Materials at Nanjing University for the support. G.Y. acknowledges the financial support from the National Key R&D Program of China (Nos. 2024YFB3715400, Nos. 2022YFA120470, 2021YFA1400400), the National Natural Science Foundation of China (No. 11974169), the Natural Science Foundation of Jiangsu Province (No. BK20233001), and the support from Nanjing University International Collaboration Initiative. L.W. acknowledges the National Key Projects for Research and Development of China (Nos. 2022YFA120470, 2021YFA1400400), National Natural Science Foundation of China (No. 12074173), and Natural Science Foundation of Jiangsu Province (Nos. BK20220066 and BK20233001). R.D. acknowledges the grant from the National Natural Science Foundation of China (No. 12004173).